\documentclass[twocolumn,epsfig,amsfonts,showpacs,prl]{revtex4}
\usepackage{epsfig}

\begin{document}

\title{Quantum oscillations of self-dual Abrikosov-Nielsen-Olesen vortices}

\author{A. Alonso Izquierdo$^{(1)}$,
W. Garcia Fuertes$^{(2)}$, M. de la Torre Mayado$^{(3)}$ and J.
Mateos Guilarte$^{(4)}$}

\affiliation{$^{(1)}$ Departamento de Matematica
Aplicada, Universidad de Salamanca, SPAIN \\
$^{(3)}$ Departamento de Fisica, Universidad de Salamanca, SPAIN
\\ $^{(4)}$ Departamento de Fisica and IUFFYM, Universidad de Salamanca, SPAIN \\
$^{(2)}$ Departamento de Fisica, Universidad de Oviedo, SPAIN}

\begin{abstract}
The mass shift induced by one-loop quantum fluctuations on
self-dual ANO vortices is computed using heat kernel/generalized
zeta function regularization methods. The quantum masses of
super-imposed multi-vortices with vorticity lower than five are
given. The case of two separate vortices with a quantum of
magnetic flux is also discussed.
\end{abstract}


\maketitle

\section{Introduction}

In this paper we present some new results on the quantization of
the self dual multi-vortex solutions of the Abelian Higgs model.
We also take the opportunity to offer a detailed description of
the concepts and techniques that allowed us to compute the
one-loop quantum correction to the mass of self-dual
Abrikosov-Nielsen-Olesen vortices with one quantum of magnetic
flux in the Rapid Communication \cite{AGMT}. The AHM  provides a
theoretical ground in several fields of physics: it provides shape
to interesting truncations of the electroweak or grand unified
theories, it also provides the basis for the various
phenomenological models for cosmic strings, or it can be used as a
Ginzburg-Landau theory for superconductivity.

Interest in this research, developed in the supersymmetric
framework in \cite{sc},\cite{le}, was rekindled two years ago.
Non-vanishing quantum corrections to the mass of $N=2$
supersymmetric vortices were recently reported in papers
\cite{Vass} and \cite{Reb}, see also {\footnote{In \cite{Vass}
there is the following footnote: \lq\lq Since no analytic form for
the profile functions of the ANO vortices is available,
calculations of the mass shift in a non-supersymmetric case is a
rather complicated problem".}}. In the second paper, it was found
that the central charge of the $N=2$ SUSY algebra receives a
non-vanishing one-loop correction that is exactly equal to the
one-loop mass shift; thus, one could talk in terms of one-loop BPS
saturation. This result fits in a pattern first conjectured in
\cite{Reb1} and then proved in \cite{Shif} for supersymmetric
kinks. Another work by the authors of the Stony Brook/Viena group,
\cite{Reb2} unveils a similar kind of behavior of supersymmetric
BPS monopoles in $N=2$ SUSY Yang-Mills theory. In this reference,
however, it is pointed out that (2+1)-dimensional SUSY vortices do
not behave exactly in the same way as their (1+1)- and
(3+1)-dimensional cousins. One-loop corrections in the vortex case
are in no way related to an anomaly in the conformal central
charge, contrarily to the quantum corrections for SUSY kinks and
monopoles.

We shall focus, however, on the purely bosonic Abelian Higgs model
and rely on the heat kernel/generalized zeta function
regularization method that we developed in reference \cite{Aai1}.
Our approach profits from the high-temperature expansion of the
heat function, which is compatible with Dirichlet boundary
conditions in purely bosonic theories. In contrast, the
application of a similar regularization method to the
supersymmetric kink requires SUSY-friendly boundary conditions,
see \cite{Vass1}. In \cite{Aai1} the kink quantum correction in
the $\phi^4$ model is estimated by this method and compared with
the correct answer obtained from the Dashen-Hasslacher-Neveu
formula, \cite{DHN} in order to check the reliability of our
approach. The relative error found is approximately 0.07\%. In
\cite{Aai2} and \cite{Aai3} we also calculated the quantum mass
corrections for kinks arising in two-component scalar models,
where second-order small fluctuations are ruled by matrix
differential operators. Therefore, we were led to generalize the
zeta function method to the matrix case, because the DHN approach,
based on a direct computation of the spectral density, is not
efficient for matrix differential Schrodinger operators. This step
has proved to be crucial, opening the possibility of applying our
method to two-dimensional topological defects in the Abelian Higgs
model.

In order to accomplish this task we shall encounter more
difficulties than for one-dimensional multi-component kinks. As
noticed by Vassilevich, the lack of analytical expressions for
vortex solutions forces us to perform a numerical analysis already
at the classical level to solve the field equations. Also, the
high-temperature expansion of the heat trace becomes more involved
due to the jump from one to two spatial dimensions;  the
recurrence relations hold between partial -rather than ordinary-
derivatives of the high-T expansion coefficients. We stress that
the evaluation of the Seeley coefficients is a very laborious
task: fluctuations of the vector, Higgs and Goldstone fields are
governed by one $4\times 4$-matrix differential operator, whereas
fluctuations of the ghosts are determined by one scalar
differential operator acting on $L^2({\Bbb R}^2)$. There is,
however, one point where the situation is more favorable as
compared to the kink case: the generalized zeta function
regularization method provides us directly with a finite quantity,
without the need of infinite renormalizations. This fact is
peculiar to even spatial dimensions and is probably related to the
lack of anomalies when fermions are added. As for kinks, we shall
obtain a simple formula for the one-loop quantum mass correction
depending on the Seeley coefficients and the number of zero modes.

One remarkable aspect of our results is that the correction found
by this method in the bosonic system is essentially twice the
correction arising in the supersymmetric case in \cite{Vass} and
\cite{Reb}. This seems to be in agreement with the relationship
between the supersymmetric and non-supersymmetric one-loop
corrections to the masses of the sine-Gordon and $\phi^4$ kinks,
see \cite{Schf} and \cite{Wimm}.

The organization of the paper is as follows: In Section \S.2 we
revise the perturbative sector of the Abelian Higgs model in the
Feynman-'t Hooft renormalizable gauge and set the one-loop mass
renormalization conventions. Section \S.3 is devoted to studying
ANO vortex solutions and their fluctuations in a partially
analytical, partially numerical manner. The high-temperature
expansion of the pertinent heat traces is developed in Section
\S.4 . Section \S.5 explains how quantum oscillations of vortices
are accounted for in the framework of generalized zeta function
regularization. In Section \S.6 the one-loop vortex mass shift
formula is applied to cylindrically symmetric self-dual vortices.
We also briefly discuss how the shift depends on the distance
between centers of a two-vortex solution. Finally, we offer a
Summary and Outlook.

\section{The planar Abelian Higgs model}

\subsection{The model}

The AHM  describes the minimal coupling between an $U(1)$-gauge
field and a scalar field in a phase where the gauge symmetry is
spontaneously broken. Defining non-dimensional space-time
variables, $x^\mu \rightarrow \frac{1}{ev} x^\mu$, and fields,
$\phi \rightarrow v\phi=v(\phi_1+i\phi_2)$, $A_\mu \rightarrow v
A_\mu$, from the vacuum expectation value of the Higgs field $v$
and the $U(1)$-gauge coupling constant $e$, the action for the
Abelian Higgs model in (2+1)-dimensions reads:
\[
S= \frac{v}{e}\int d^3 x \left[ -\frac{1}{4} F_{\mu \nu} F^{\mu
\nu}+\frac{1}{2} (D_\mu \phi)^* D^\mu \phi - U(\phi,\phi^*)
\right]
\]
with
\[
U(\phi,\phi^*)=\frac{\kappa^2}{8} (\phi^* \phi-1)^2
\hspace{0.5cm} .
\]
$\kappa^2=\frac{\lambda}{e^2}$ is the only classically relevant
parameter and measures the ratio between the square of the masses
of the Higgs, $M^2=\lambda v^2$, and vector particles,
$m^2=e^2v^2$; $\lambda$ is the Higgs field self-coupling. We
choose a system of units where $c=1$, but $\hbar$ has dimensions
of length $\times$ mass. Also, we define the metric tensor as:
$g_{\mu\nu}={\rm diag}(1,-1,-1), \, \mu,\nu=0,1,2$.

\subsection{Feynman rules in the R-gauge}

The choice of $\phi^V=1$ as the ground state causes spontaneous
symmetry breaking of the Abelian gauge invariance. In the
Feynman-'t Hooft renormalizable gauge,
\[
R(A_\mu , G)=\partial_\mu A^\mu - G \qquad \qquad ,
\]
the particle spectrum involves a vector particle $A_\mu$, Higgs
and Goldstone scalar particles $\phi= 1+H+iG$, and a complex ghost
$\chi$. The Feynman rules are read from the action, see Reference
\cite{Velt}:
\begin{eqnarray*}
S&=&{v\over e}\int \, d^3x \, \left[ -\frac{1}{2} A_\mu
[-g^{\mu\nu}(\partial_\alpha\partial^\alpha +1)]A_\nu \right.\\
&+&\frac{1}{2}\partial_\mu G\partial^\mu G-\frac{1}{2} G^2+
\frac{1}{2}\partial_\mu H\partial^\mu H-\frac{\kappa^2}{2}  H^2\\
&+&\partial_\mu\chi^*\partial^\mu \chi- \chi^*\chi -{\kappa^2\over
2}H (H^2+G^2)\\&+& A_\mu (\partial^\mu H
G-\partial^\mu G H)+H (A_\mu A^\mu -\chi^* \chi)\\
&-&\left. \frac{\kappa^2}{8} (H^2+G^2)^2 +\frac{1}{2}(G^2+H^2)
A_\mu A^\mu \right] \qquad .
\end{eqnarray*}
There are four propagators, plus five third-order and five
fourth-order vertices shown in the next two Tables:

\begin{table}[h]
\begin{center}
\caption{Propagators}
\begin{tabular}{lccc} \\ \hline
\textit{Particle} & \textit{Field} & \textit{Propagator} & \textit{Diagram} \\
\hline \\ Higgs & $H(x)$ & $\displaystyle\frac{i e \hbar
}{v(k^2-\kappa^2+i\varepsilon)}$ &
\parbox{2.5cm}{\epsfig{file=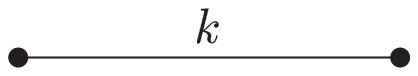,width=2.25cm}} \\[0.5cm]
Goldstone & $G(x)$ & $\displaystyle\frac{ie
\hbar}{v(k^2-1+i\varepsilon)}$ &
\parbox{2.5cm}{\epsfig{file=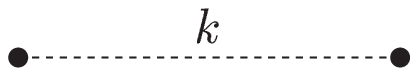,width=2.25cm}} \\[0.5cm]
Ghost & $\chi$ & $\displaystyle\frac{ie
\hbar}{v(k^2-1+i\varepsilon)}$ &
\parbox{2.5cm}{\epsfig{file=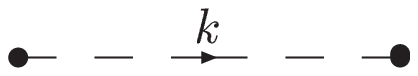,width=2.25cm}} \\[0.5cm]
Vector Boson & $A_\mu(x)$  & $\frac{-ie \hbar g^{\mu\nu}
}{v(k^2-1+i\varepsilon)}$ &
\parbox{2.5cm}{\epsfig{file=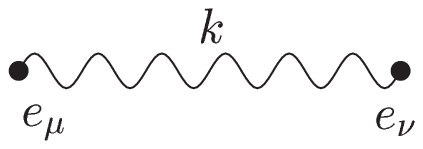,width=2.1cm}}
\\[0.5cm] \hline
\end{tabular}
\end{center}
\end{table}

\begin{table}[hbt]
\begin{center}
\caption{Third- and fourth-order vertices }
\begin{tabular}{clcl} \\ \hline
\textit{Vertex} & \textit{Weight} & \textit{Vertex} & \textit{Weight} \\
\hline \\
\parbox{2.3cm}{\epsfig{file=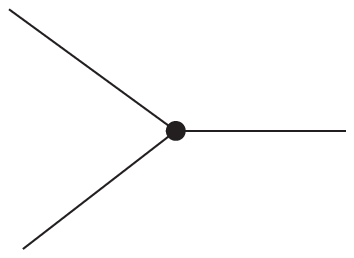,width=1.8cm}} &
$\displaystyle -3i\kappa^2\frac{v}{\hbar e} $  &
\parbox{2.3cm}{\epsfig{file=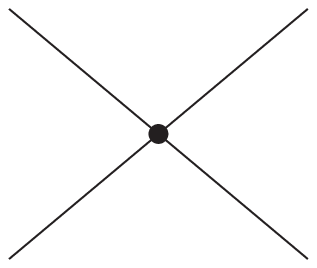,width=1.8cm}} &
$\displaystyle -3i\kappa^2\frac{v}{\hbar e}$ \\[0.5cm]
\parbox{2.3cm}{\epsfig{file=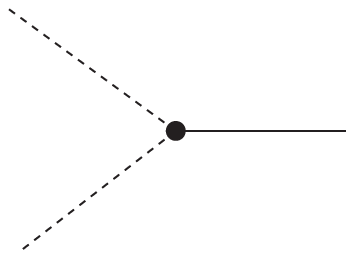,width=1.8cm}} &
$\displaystyle -i\kappa^2\frac{v}{\hbar e}$  &
\parbox{2.3cm}{\epsfig{file=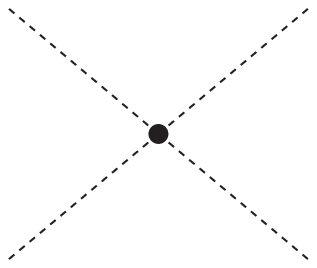,width=1.8cm}} &
$\displaystyle -3i\kappa^2\frac{v}{\hbar e}$ \\[0.5cm]
 \parbox{2.3cm}{\epsfig{file=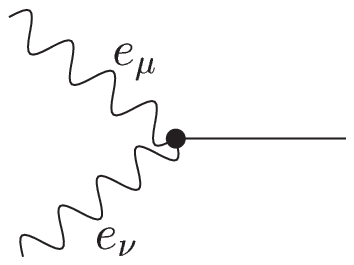,width=1.8cm}} &
$\displaystyle 2i\frac{v}{\hbar e} g^{\mu \nu}$  &
\parbox{2.3cm}{\epsfig{file=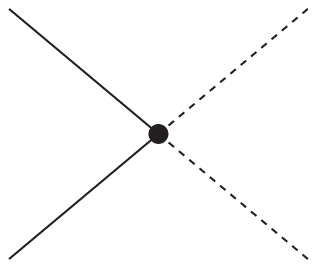,width=1.8cm}} &
$\displaystyle -i\kappa^2\frac{v}{\hbar e}$
\\[0.5cm]
 \parbox{2.3cm}{\epsfig{file=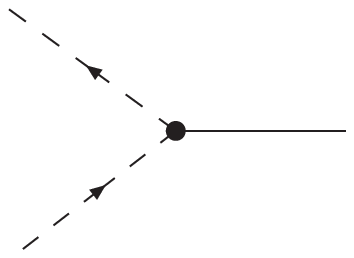,width=1.8cm}} &
$\displaystyle -i\frac{v}{\hbar e}$ &
\parbox{2.3cm}{\epsfig{file=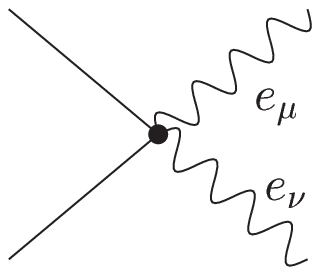,width=1.8cm}} &
$\displaystyle 2 i\frac{v}{\hbar e}g^{\mu \nu}$\\[0.5cm]
\parbox{2.3cm}{\epsfig{file=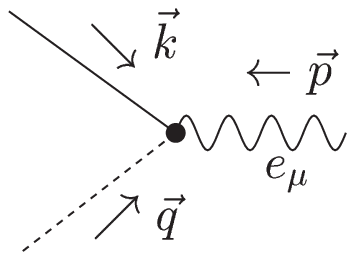,width=1.8cm}} &
$\displaystyle (k^\mu-q^\mu)\frac{v}{\hbar e}$  &
\parbox{2.3cm}{\epsfig{file=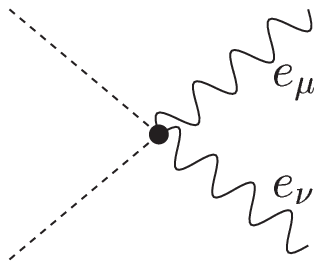,width=1.8cm}} &
$\displaystyle 2 i \frac{v}{\hbar e} g^{\mu \nu}$ \\[0.7cm] \hline
\end{tabular}
\end{center}
\end{table}

\subsection{One-loop renormalization}

Defining
\[
I(c^2)=\int \, \frac{d^3k}{(2\pi)^3} \cdot
\frac{i}{k^2-c^2+i\varepsilon}
\]
and bearing in mind that $I(\kappa^2)=I(1)+\mbox{finite part}$,
the one-loop divergences in the planar Abelian Higgs model can be
organized as follows:
\begin{itemize}

\item Higgs tadpole

\centerline{\parbox{1.5cm}{\epsfig{file=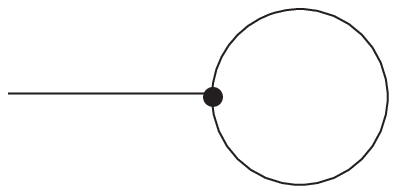,width=1.5cm}}
$+$
\parbox{1.5cm}{\epsfig{file=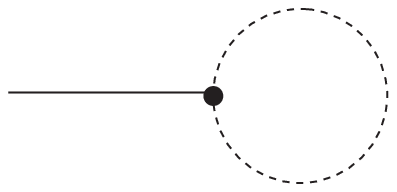,width=1.5cm}} $+$
\parbox{1.5cm}{\epsfig{file=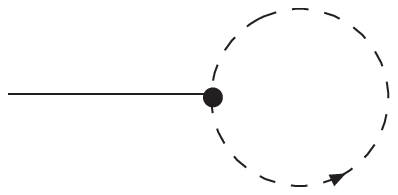,width=1.5cm}} $+$
\parbox{1.5cm}{\epsfig{file=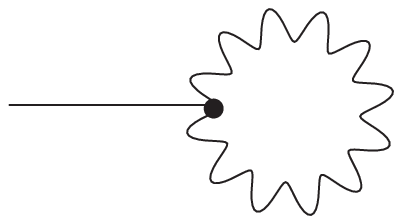,width=1.5cm}} $=$}
\[
= - 2 i ( \kappa^2 + 1) I(1) + \mbox{finite part}
\]

\item Higgs propagator

\centerline{\parbox{1.5cm}{\epsfig{file=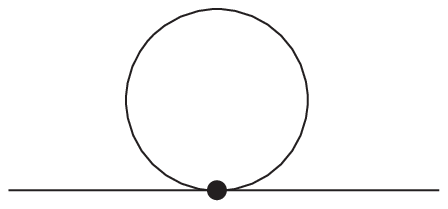,width=1.5cm}}
$+$
\parbox{1.5cm}{\epsfig{file=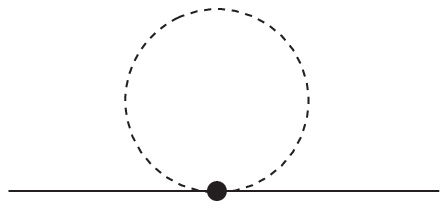,width=1.5cm}} $+$
\parbox{1.5cm}{\epsfig{file=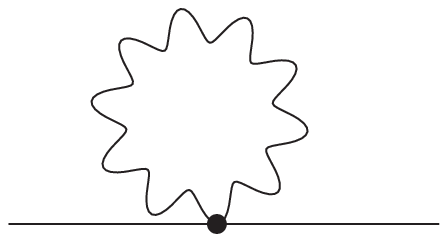,width=1.5cm}} $+$
\parbox{1.5cm}{\epsfig{file=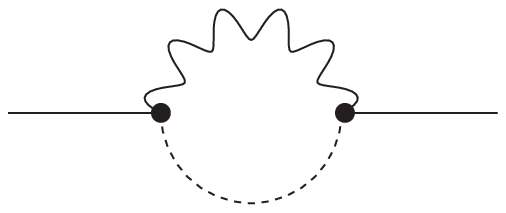,width=1.5cm}} $=$}
\[
= - 2 i ( \kappa^2+ 1) I(1) + \mbox{finite part}
\]

\item Goldstone propagator

\centerline{\parbox{1.5cm}{\epsfig{file=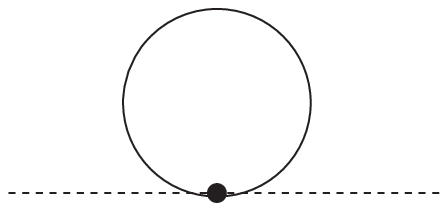,width=1.5cm}}
$+$
\parbox{1.5cm}{\epsfig{file=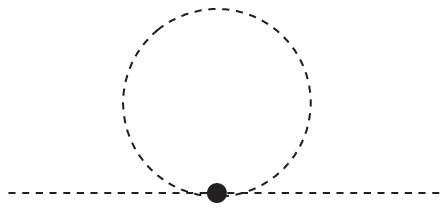,width=1.5cm}} $+$
\parbox{1.5cm}{\epsfig{file=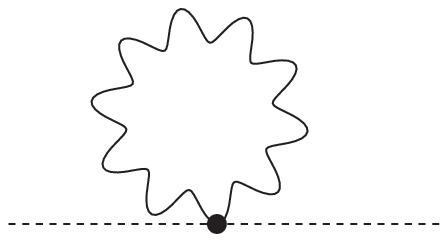,width=1.5cm}} $+$
\parbox{1.5cm}{\epsfig{file=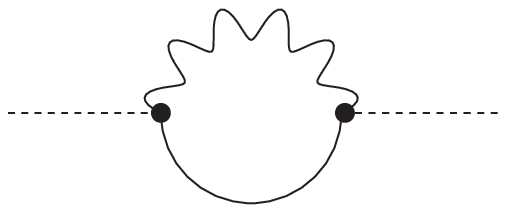,width=1.5cm}} $=$}
\[
= - 2 i ( \kappa^2+ 1) I(1) + \mbox{finite part}
\]

\item Vector boson propagator

\centerline{\parbox{1.5cm}{\epsfig{file=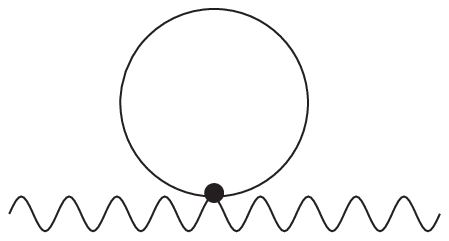,width=1.5cm}}
$+$
\parbox{1.5cm}{\epsfig{file=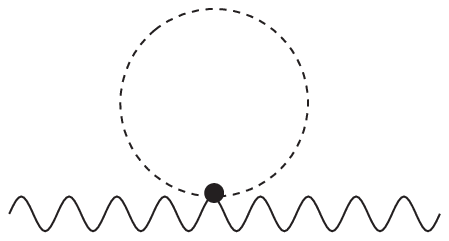,width=1.5cm}} $+$
\parbox{1.5cm}{\epsfig{file=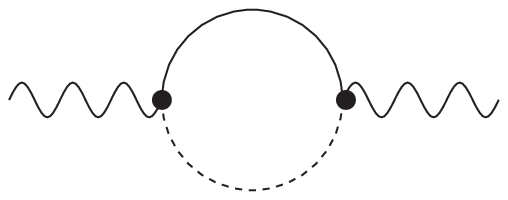,width=1.5cm}} $=$}
\[
=  2 i I(1) + \mbox{finite part}
\]

\end{itemize}
There are no more one-loop divergent graphs. Therefore, in a
minimal subtraction scheme, we add the diagrams shown in the next
Table to cancel the divergences in the one-loop graphs.

\begin{table}[hbt]
\begin{center}
\caption{One-loop counter-terms }
\begin{tabular}{ccc} \\ \hline
\textit{Diagram} & \hspace{0.3cm} & \textit{Weight} \\
\hline
\parbox{2.3cm}{\epsfig{file=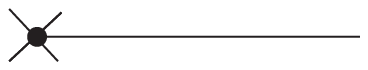,width=2.3cm}} & &
$\displaystyle 2i(\kappa^2+1)I(1) $
\\
\parbox{2.3cm}{\epsfig{file=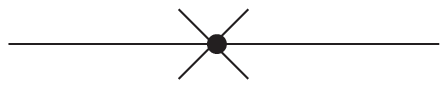,width=2.3cm}} & &
$\displaystyle 2i(\kappa^2+1) I(1)$ \\
 \parbox{2.3cm}{\epsfig{file=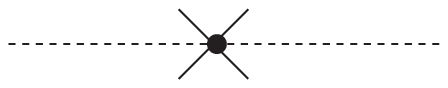,width=2.3cm}} & &
$\displaystyle 2i(\kappa^2+1) I(1)$
\\
 \parbox{2.3cm}{\epsfig{file=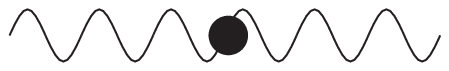,width=2.3cm}} & &
$\displaystyle -2i I(1)$ \\[0.5cm] \hline \\[0.25cm]
\end{tabular}
\end{center}
\end{table}
This is tantamount to considering that the counter-terms
\begin{eqnarray}
{\cal L}_{c.t.}^S &=& \frac{\hbar(\kappa^2+1)}{2} I(1)
\left[|\phi|^2-1 \right]  \label{eq:ct1}\\
{\cal L}_{c.t.}^A &=& -{\hbar \over 2}I(1) A_\mu A^\mu
\label{eq:ct2}
\end{eqnarray}
enter into the Lagrangian.

All the finite parts are proportional to $I(\kappa^2)-I(1)$ and
they vanish in the critical point between Type I and Type II
superconductivity, $\kappa^2=1$, to be considered in the sequel.
Note that the mass of the elementary particles for this critical
value of $\kappa$ is taken as subtraction point so that the
counter-terms exactly cancel the divergence due to the Higgs
tadpole. Therefore, our renormalization criterion is equivalent to
the renormalization condition stipulated in \cite{Vass} and
\cite{Reb} when $\kappa^2=1$.

\section{ANO self-dual vortices}

Abrikosov-Nielsen-Olesen  vortices are topological defects
satisfying the time-independent field equations:
\begin{equation}
\partial_i F_{ij} = J_j \hspace{0.3cm};\hspace{0.2cm}
\frac{1}{2} D_i D_i \phi= \frac{\partial U}{\partial \phi^*}
\qquad , \label{eq:soe}
\end{equation}
where $J_j= \frac{i}{2} \left(  \phi^* D_j \phi - (D_j \phi)^* \,
\phi \right)$ is the electric current. They are static and
localized solutions for which the energy
\begin{equation}
E= \int d^2 x [\frac{1}{4} F_{ij} F_{ij} + \frac{1}{2} (D_i
\phi)^* D_i \phi + \frac{\kappa}{8} (\phi^* \phi-1)^2 ]
\label{eq:he}
\end{equation}
is finite. Thus, ANO vortices comply with the boundary conditions
on $S_\infty^1$, i.e. when $r=\sqrt{x_1^2+x_2^2}$ tends to
$\infty$:
\begin{equation}
\left. \phi^{*} \phi \right|_{S_\infty^1} =1 \hspace{0.2cm} ,
\hspace{0.2cm} D_i \phi |_{S_\infty^1} =(\partial_i \phi - i A_i
\phi)|_{S_\infty^1}=0 \hspace{0.2cm} , \label{eq:bc}
\end{equation}
i.e., $\phi |_{S_\infty^1} = e^{il \theta}$, $l\in{\mathbb Z}$,
and $A_i |_{S_\infty^1} =- i \phi^*\partial_i \phi|_{S_\infty^1}$.

\subsection{First-order equations}

For the value of the coupling constant $\kappa^2=1$, the energy
functional can be arranged as follows
\[
E= \int \frac{d^2 x}{2} \left( |D_1 \phi \pm i D_2 \phi|^2 + [
F_{12} \pm {\textstyle\frac{1}{2}} (\phi^* \phi-1) ]^2
\right)+\frac{1}{2}|g|
\]
where $g= \int d^2 x F_{12}=2{\pi l}$ is the non-dimensional
quantized magnetic flux. Solutions satisfying the first-order
differential equations
\[
D_1 \phi \pm i D_2 \phi=0 \hspace{0.3cm};\hspace{0.3cm} F_{12} \pm
\frac{1}{2} (\phi^*\phi-1) =0
\]
or, equivalently,
\begin{eqnarray}
&&(\partial_1 \phi_1 +A_1 \phi_2)\mp(\partial_2 \phi_2-A_2
\phi_1)=0  \label{eq:foe1} \\
&&\pm(\partial_2 \phi_1 +A_2 \phi_2)+(\partial_1 \phi_2-A_1
\phi_1)=0 \label{eq:foe2}\\
&&F_{12}\pm\frac{1}{2}(\phi_1^2+\phi_2^2-1)=0
\label{eq:foe3}\qquad .
\end{eqnarray}
also solve the second-order equations (\ref{eq:soe}) and are
called ANO self-dual vortices if they also satisfy the boundary
conditions (\ref{eq:bc}). In what follows, we shall focus on
solutions with positive $l$: i.e., we shall choose the upper signs
in the first-order equations.

\subsection{Self-dual vortices with cylindrical symmetry}

If $\theta={\rm arctan}\frac{x_2}{x_1}$ is the polar angle, the
ansatz
\begin{eqnarray*}
\phi_1(x_1,x_2) = f(r) {\rm cos}l\theta \quad &,& \quad
\phi_2(x_1,x_2) = f(r) {\rm sin}l\theta \\
A_1(x_1,x_2) =-l \frac{\alpha(r)}{r}{\rm sin}\theta \quad &,&
\quad A_1(x_1,x_2) = l \frac{\alpha(r)}{r}{\rm cos}\theta
\end{eqnarray*}
plugged into the first-order equations (\ref{eq:foe1},
\ref{eq:foe2}, \ref{eq:foe3}) leads to:
\begin{equation}
{1\over r} {d \alpha \over d r}= \mp \frac{1}{2 l} (f^2-1) \quad ,
\quad {d f\over d r}  = \pm \frac{l}{r} f(r)[1-\alpha(r)] \quad .
\label{eq:rrfo}
\end{equation}
Regular solutions of (\ref{eq:rrfo}) with the boundary conditions
${\displaystyle \lim_{r\rightarrow\infty}} f(r) = 1$,
${\displaystyle \lim_{r\rightarrow\infty}} \alpha(r) = 1$, zeroes
of the Higgs and vector fields at the origin, $f(0) =0$,
$\alpha(0)=0$, and integer magnetic flux,
\[
g= - \oint_{r=\infty} dx_i A_i = -l\oint_{r=\infty}{
[x_2dx_1-x_1dx_2]\over r^2}=2 \pi l \,\, ,
\]
exist and can be found by a mixture of analytical and numerical
methods.

Following the procedure developed in \cite{VeSh}, we obtain
numerical solutions for the vortex equations (\ref{eq:rrfo}).
Indeed, this approach gives the vortex solution in three different
ranges of the radial coordinate. For small values of $r$, a power
series is tested in the first-order differential equations
(\ref{eq:rrfo}), leading to a recurrence relation between the
coefficients. Reference \cite{VeSh} also describes the asymptotic
behavior of the solutions. Thus, a numerical scheme can be
implemented by setting a boundary condition in a non-singular
point of (\ref{eq:rrfo}), which is obtained from the power series
for small values of $r$. This numerical method provides us with
the behavior of the vortex solutions for intermediate distances by
means of an interpolating polynomial which passes through the
numerical data.

The results are shown in figure 1, where the field profiles
$\alpha(r)$ and $f(r)$, the magnetic field $B(r)={l\over
2r}\frac{d\alpha}{dr}$ and the energy density
\[
\varepsilon(r)={1\over 4}(1-f^2(r))^2+{l^2\over
r^2}(1-\alpha(r))^2 f^2(r)
\]
are plotted with respect to $r$ for self-dual ANO vortices with
$l=1$, $l=2$, $l=3$, and $l=4$. A three-dimensional view of the
energy density in the plane is also shown in figure 2 for $l=1$,
$l=2$, $l=3$, and $l=4$ self-dual vortices. Note that the $l=1$
vortex shows a different pattern as compared with flux tubes of
several quanta: only in the first case is the energy density
maximum at the origin (the center).
\begin{widetext}
\begin{center}
\includegraphics[height=2.cm]{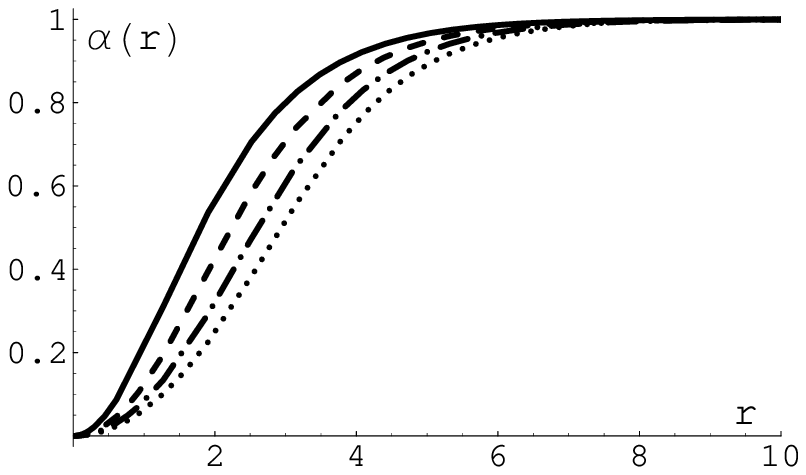
}\hspace{0.6cm}
\includegraphics[height=2.cm]{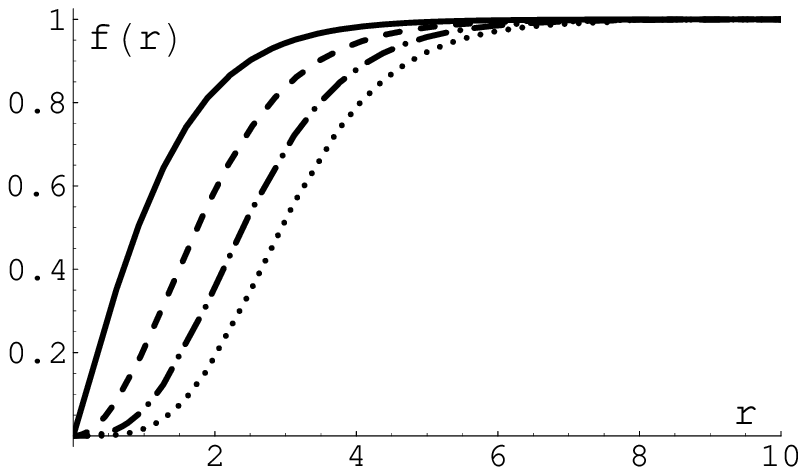}\hspace{0.6cm}
\includegraphics[height=2.cm]{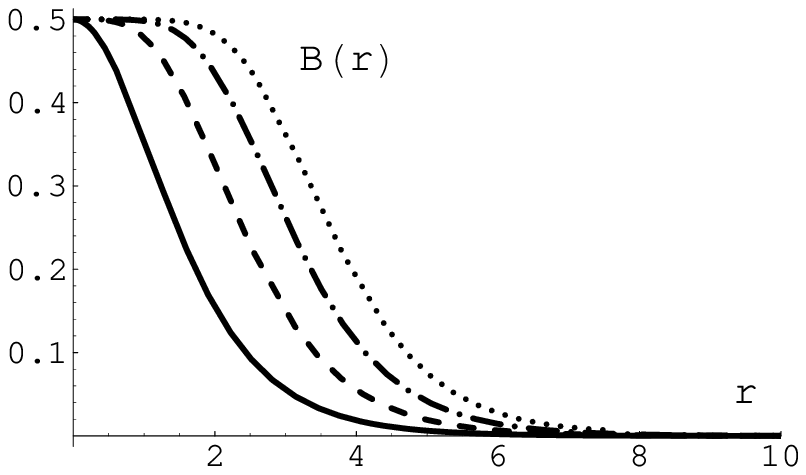}\hspace{0.6cm}
\includegraphics[height=2.cm]{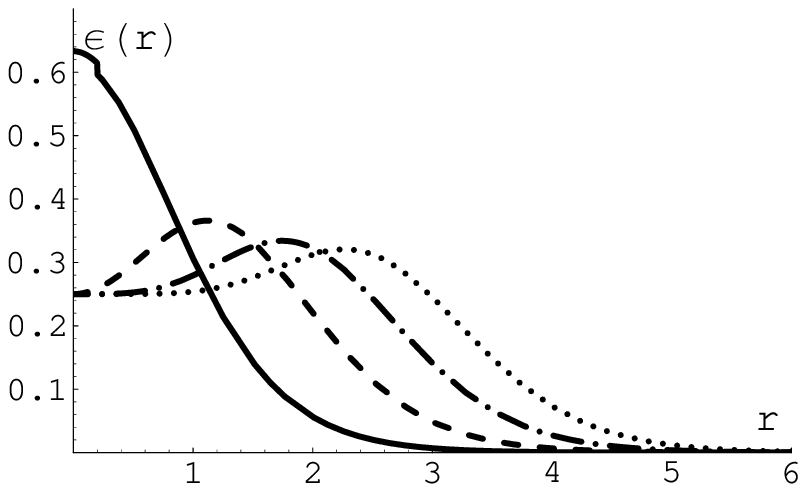} \\
{\small Figure 1. \textit{Plots of the field profiles $\alpha(r)$
(a) and $f(r)$ (b), the magnetic field $B(r)$ (c), and the energy
density $\varepsilon(r)$ for vortices with $l=1$ (solid line),
$l=2$ (broken line), $l=3$ (broken-doted line) and $l=4$ (doted
line).}}
\end{center}
\begin{center}
\includegraphics[height=2.6cm]{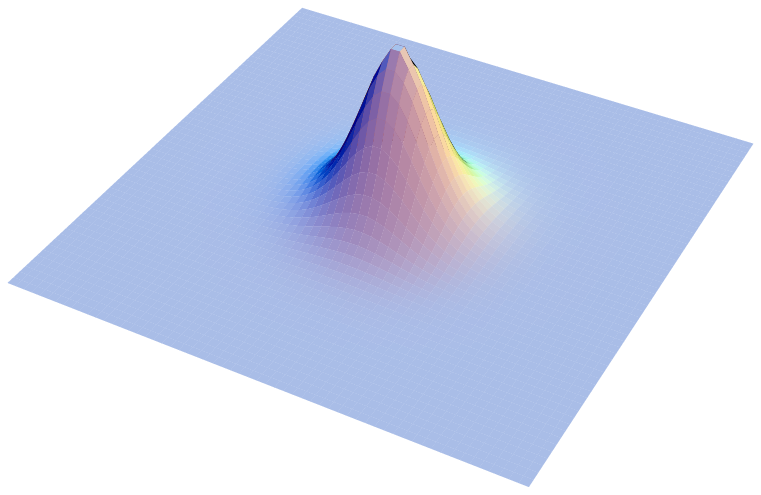}\hspace{0.6cm}
\includegraphics[height=2.6cm]{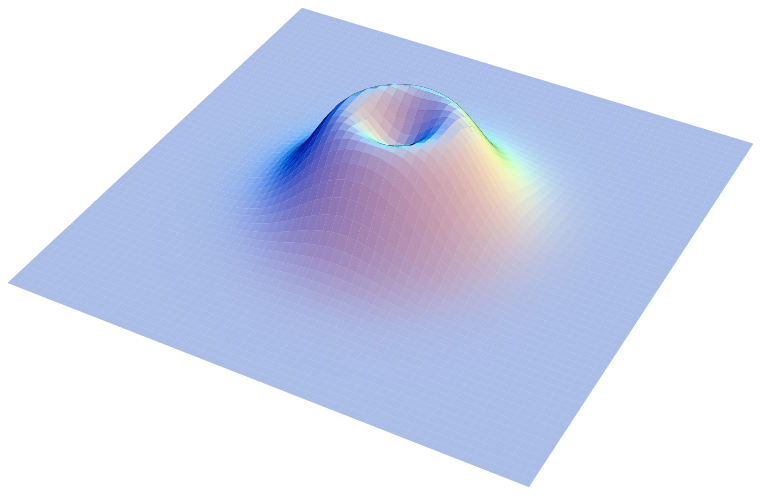}\hspace{0.6cm}
\includegraphics[height=2.6cm]{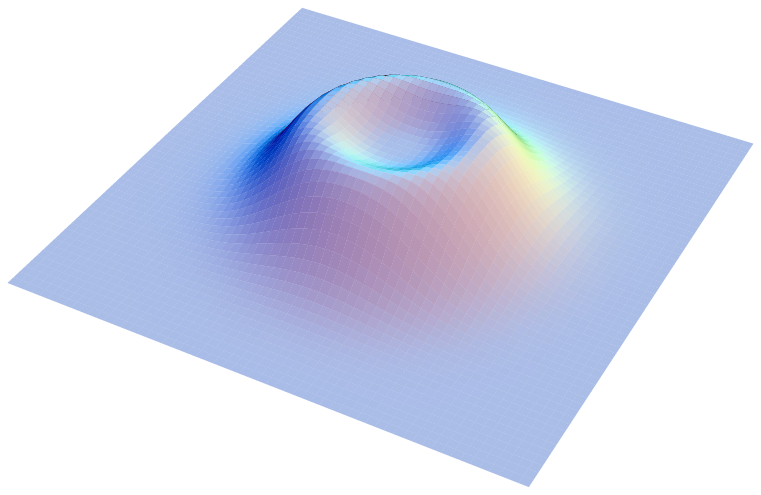}\hspace{0.6cm}
\includegraphics[height=2.6cm]{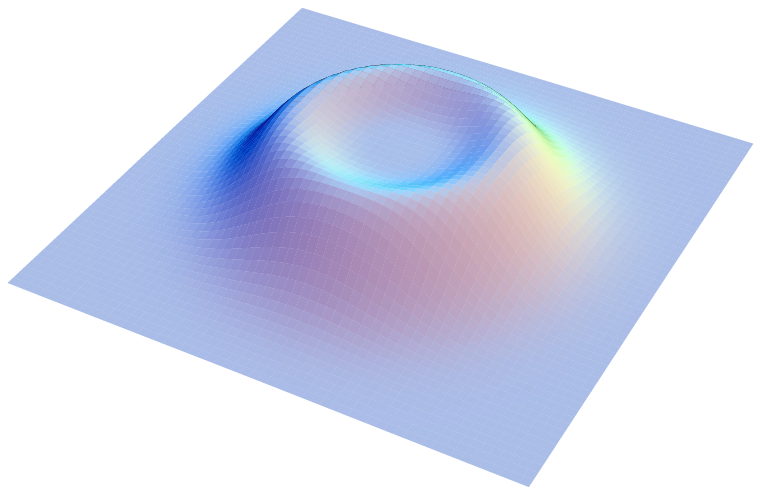} \\
{\small Figure 2. \textit{3D graphics of the energy density for
$l=1$, $l=2$, $l=3$ and $l=4$ self-dual symmetric ANO vortices.}}
\end{center}
\end{widetext}

\subsection{Two-vortex solutions with distinct centers}

To tackle the task of building $l=2$ ANO self-dual solutions
formed by two $l=1$ vortices with centers separated by a distance
$d$, we follow the work \cite{JR} by Jacobs and Rebbi. A
variational method is implemented in two stages:

In the first stage, trial functions depending only on a single
variational parameter $w$ are considered:
\begin{widetext}
\begin{eqnarray}
\phi_\omega (z,z^*)&=& \Phi(z,z^*) \, \left[ \omega \,
f^{(1)}(|z-d/2|)\, f^{(1)}(|z+d/2|)+ (1-\omega)\,
\frac{|z^2-(d/2)^2|}{|z^2|}\, f^{(2)}(|z|)\right] \label{eq:fun1} \\
A^\omega (z,z^*) &=& \omega \left( \frac{i}{z^*-d/2}\,
\alpha^{(1)}(|z-d/2|)+ \frac{i}{z^*+d/2}\,
\alpha^{(1)}(|z+d/2|)\right)+(1-\omega)\, \frac{2 i}{z^*} \,
\alpha^{(2)}(|z|) \label{eq:fun2} \qquad .
\end{eqnarray}
\end{widetext}
Here
\[
z=x_1+i x_2 \qquad , \qquad
A^\omega(z,z^*)=A_1^\omega(z,z^*)+iA_2^\omega(z,z^*) \qquad ,
\]
and
\[
\Phi=\sqrt{\frac{z^2-(d/2)^2}{z^{*2}-(d/2)^2}}
\]
is essentially a phase chosen in such a way that the magnetic flux
is equal to $4\pi$. $f^{(1)}$, $\alpha^{(1)}$, $f^{(2)}$ and
$\alpha^{(2)}$ stand for the functions $f$ and $\alpha$ associated
with self-dual solutions with cylindrical symmetry -obtained in
the previous subsection- respectively with vorticity $l=1$ and
$l=2$. Evoking (\ref{eq:fun1}) and (\ref{eq:fun2}) we expect that
$\omega=0$ for the case $d=0$ and $\omega=1$ for the case $d>>1$.
Plugging (\ref{eq:fun1}) and (\ref{eq:fun2}) into the energy
functional, we obtain a expression $E(\omega )$, which is set to
be minimized as a function of $\omega$.

In the second stage the trial functions are refined by adding a
deformation such that two requirements are fulfilled: 1) the
scalar field vanishes at the two centers. 2) the gauge-invariant
quantities associated with the solution are symmetric with respect
to the reflection $z\rightarrow z^*$. The invariant ansatz reads:
\begin{widetext}
\begin{eqnarray*}
\phi(z,z^*)&=& \phi_\omega(z,z^*)+\Phi(z,z^*)\left|z^2-(d/2)^2
\right| (\cosh |z|)^{-1} \sum_{i=0}^N \sum_{j=0}^i f_{ij}
\frac{(zz^*)^i}{2} \left[ \left( \frac{z}{z^*}\right)^j +\left(
\frac{z^*}{z}\right)^j \right] \\
A(z,z^*)&=& A^\omega (z,z^*)+ \frac{1}{\cosh |z|} \left\{ z
\sum_{i=0}^N \sum_{j=0}^i a_{ij}^{I} \frac{(zz^*)^i}{2} \left[
\left( \frac{z}{z^*}\right)^j +\left( \frac{z^*}{z}\right)^j
\right]+ z^* \sum_{i=0}^N \sum_{j=0}^i a_{ij}^{II}
\frac{(zz^*)^i}{2} \left[ \left( \frac{z}{z^*}\right)^j +\left(
\frac{z^*}{z}\right)^j \right] \right\}
\end{eqnarray*}
\begin{center}
\includegraphics[height=3cm]{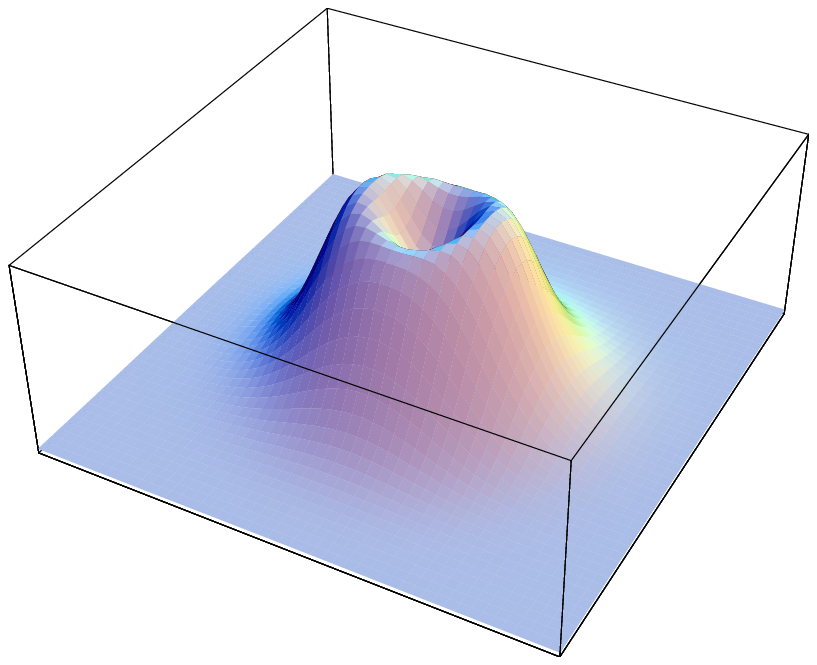}\hspace{1cm}
\includegraphics[height=3cm]{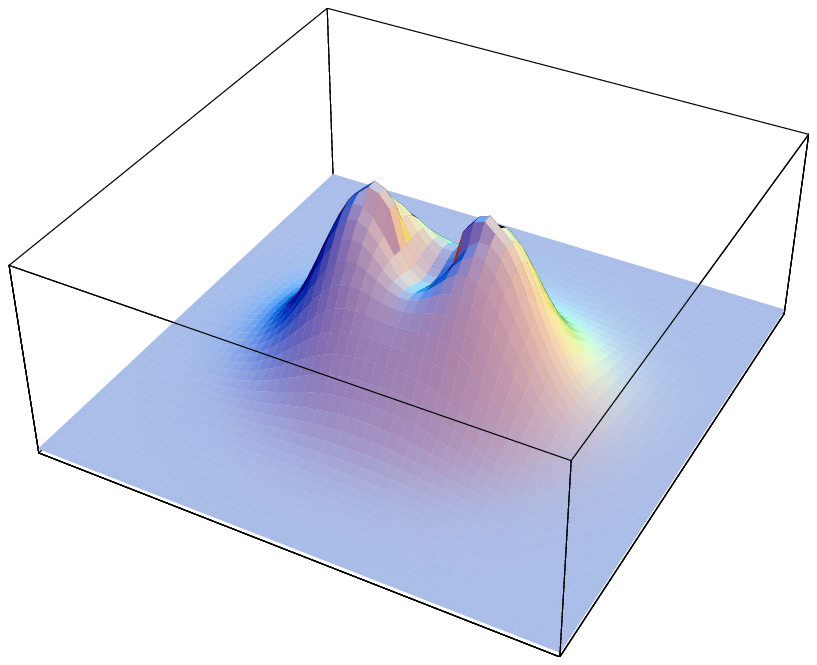}\hspace{1cm}
\includegraphics[height=3cm]{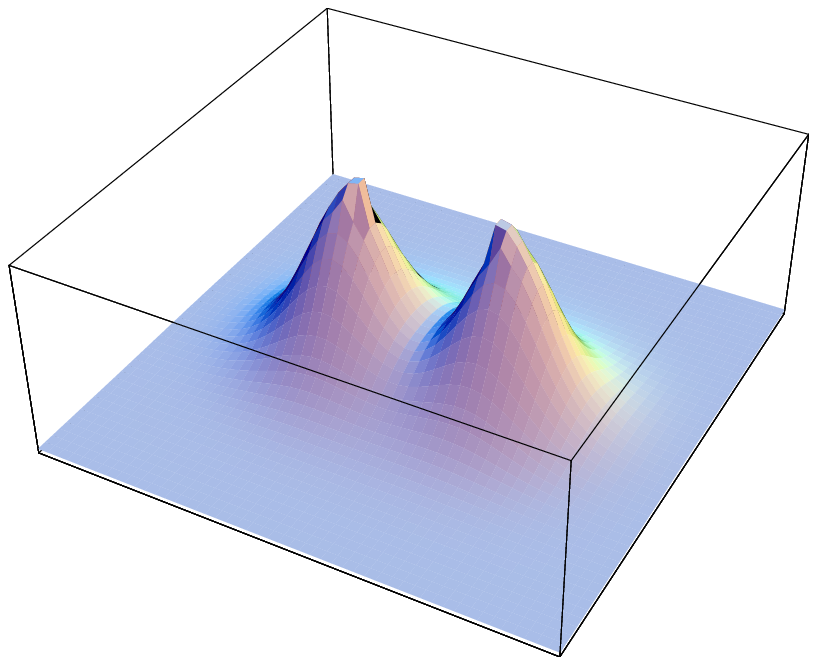}\\
{\small Figure 3. \textit{3D graphics of the energy density for
$l=2$ self-dual separate vortices with centers at distances $d=1$,
$d=2$, $d=3$.}}
\end{center}
\end{widetext}
These expressions involve $\aleph=3\frac{(N+1)(N+2)}{2}$
variational parameters $f_{ij}$, $a^I_{ij}$, $a^{II}_{ij}$.
Finding the minimum of the energy functional as a function of
these $\aleph$ variables - a task for Mathematica- a good
approximation to the $l=2$ self-dual solution with a distance $d$
between the two $l=1$ vortex centers is obtained. For our purposes
setting $N=1$ such that $\aleph=9$ will suffice. The energy
density for two-vortex solutions found by this method if
$\aleph=9$ is depicted for $d=1$, $d=2$ and $d=3$ in the above
figure.

\subsection{Small fluctuations}

We generically denote the vortex solution fields as
\[
\phi^V=\psi=\psi_1+i\psi_2 \qquad , \qquad A_k^V=V_k \qquad ,
k=1,2 \qquad .
\]
Assembling the small fluctuations around the solution
\[
\phi (\vec{x})=\psi (\vec{x})+\varphi (\vec{x}) \qquad , \qquad
A_k(\vec{x})=V_k(\vec{x})+a_k(\vec{x})
\]
in a four column $\xi(\vec{x})$, ${\rm L}^2$-integrable
second-order fluctuations around a given vortex solution are still
solutions of the first-order equations with the same magnetic flux
if they belong to the kernel of the Dirac-like operator, ${\cal
D}\xi (\vec{x})=0$, \cite{Wein}
\begin{widetext}
\[
{\cal D}\xi(\vec{x})= \left(\begin{array}{cccc} -\partial_2 &
\partial_1 & \psi_1 & \psi_2 \\ -\partial_1 & -\partial_2 &
-\psi_2 & \psi_1
\\ \psi_1 & -\psi_2 & -\partial_2+V_1 & -\partial_1-V_2 \\ \psi_2 & \psi_1 &
\partial_1+V_2 & -\partial_2+V_1 \end{array}\right)\left(\begin{array}{c} a_1(\vec{x})\\
a_2(\vec{x})\\
\varphi_1(\vec{x})\\ \varphi_2(\vec{x}) \end{array}\right)
\]
\end{widetext}
The first component of ${\cal D}\xi$ gives the deformation of the
vortex equation (\ref{eq:foe3}), whereas the third and fourth
components are due  to the respective deformation of the covariant
holomorphy equations (\ref{eq:foe2}) and (\ref{eq:foe1}). The
second component sets the background gauge
\[
B(a_k,\varphi;\psi)=\partial_k
a_k-(\psi_1\varphi_2-\psi_2\varphi_1)
\]
on the fluctuations.

The operator ${\cal H}^+={\cal D}^\dagger{\cal D}$ and its partner
${\cal H}^-={\cal D}{\cal D}^\dagger$ read:
\begin{widetext}
\begin{eqnarray*}
{\cal H}^+={\small \left(\begin{array}{cccc} -\bigtriangleup
+|\psi|^2 & 0 & -2\nabla_1\psi_2 & 2\nabla_1\psi_1
\\ 0 & -\bigtriangleup+|\psi|^2 & -2\nabla_2\psi_2 &
2\nabla_2\psi_1 \\ -2\nabla_1\psi_2 & -2\nabla_2\psi_2 &
-\bigtriangleup+{1\over 2}(3|\psi|^2+2V_kV_k-1) & -2V_k\partial_k
\\ 2\nabla_1\psi_1 & 2\nabla_2\psi_1 & 2V_k\partial_k  &
-\bigtriangleup+{1\over 2}(3|\psi|^2+2V_kV_k-1)
\end{array}\right)}
\end{eqnarray*}
\begin{eqnarray*}
{\cal H}^-={\small \left(\begin{array}{cccc}
-\bigtriangleup+|\psi|^2 & 0 & 0 & 0 \\ 0 &
-\bigtriangleup+|\psi|^2 & 0 & 0  \\ 0 & 0 &
-\bigtriangleup+{1\over 2}(|\psi|^2+1)+V_kV_k & -2V_k\partial_k
\\ 0 & 0 & 2V_k\partial_k & -\bigtriangleup+{1\over
2}(|\psi|^2+1)+V_kV_k
\end{array}\right)} \qquad .
\end{eqnarray*}
\end{widetext}
One can check that ${\cal H}^+$ arises in the small deformation of
the second-order equations (\ref{eq:soe}) in the background gauge
for $\kappa=1$, thus ruling the second-order fluctuations around
the vortex solutions. In fact, for $l=0$ one finds that ${\cal
H}^+={\cal H}^-={\cal H}_0$, where
\[
{\cal H}_0={\small \left(\begin{array}{cccc} -\bigtriangleup +1 &
0 & 0 & 0
\\ 0 & -\bigtriangleup+1 & 0 &
0 \\ 0 & 0 & -\bigtriangleup+1 & 0
\\ 0 & 0 & 0  &
-\bigtriangleup+1
\end{array}\right)}
\]
is the second-order fluctuation operator around the vacuum in the
Feynman-'t Hooft renormalizable gauge: the background gauge in the
vacuum sector. Note that the fluctuations in this gauge correspond
to a massive vector particle plus scalar Higgs and Goldstone
fields. It will be useful in the sequel to write the second-order
fluctuation operators around $l\geq 1$ vortices in the form:
\[
{\cal H}^\pm={\cal H}_0
+Q^\pm_k(\vec{x})\partial_k+V^\pm(\vec{x})\qquad ,
\]
where $Q^\pm_k(\vec{x})$ and $V^\pm(\vec{x})$ are $4\times 4$
functional matrices.

\section{High-temperature expansion of heat traces}

\subsection{Index theorem: moduli space of self-dual vortices}

One easily checks that $\dim \ker {\cal D}^\dagger=0$ because the
spectrum of ${\cal H}^-$ is definite positive. Thus, the dimension
of the moduli space of self-dual vortex solutions with magnetic
charge $l$ is the index of ${\cal D}$:
\[
{\rm ind}\,{\cal D}=\dim \ker {\cal D}- \dim \ker {\cal D}^\dagger
\qquad \qquad .
\]
We follow Weinberg \cite{Wein}, using the background instead of
the Coulomb gauge, to briefly determine ${\rm ind}\,{\cal D}$. The
spectra of the operators ${\cal H}^+$ and ${\cal H}^-$ only differ
in the number of eigen-functions belonging to their kernels. For
topological vortices, we do not expect pathologies due to
asymmetries between the spectral densities of ${\cal H}^+$ and
${\cal H}^-$, and thus ${\rm ind}\, {\cal D}={\rm Tr}\,
e^{-\beta{\cal H}^+}-{\rm Tr}\, e^{-\beta{\cal H}^-}$. For a case
in which these asymmetries are important, see the treatment of
Chern-Simons-Higgs topological and non-topological vortices given
in \cite{Wein1,GM99}.

The heat trace of a $N\times N$ matrix differential operator
\[
{\cal H}={\cal H}_0 +Q_k(\vec{x})\partial_k+V(\vec{x})
\]
-like the ${\cal H}^\pm$ operators- is defined as
\[
{\rm Tr}\,e^{-\beta{\cal H}}={\rm tr}\int_{{\mathbb R}^2} \,
d^2\vec{x} \, K_{{\cal H}}(\vec{x},\vec{x};\beta)
\]
where $K_{{\cal H}}(\vec{x},\vec{y};\beta)$ is the $N\times N$
matrix kernel of the heat equation and ${\rm tr}$ is the usual
matrix trace. Therefore, $K_{{\cal H}}(\vec{x},\vec{y};\beta)$
solves the heat equation
\begin{equation}
\left(\frac{\partial}{\partial\beta}{\mathbb I}+{\cal H}
\right)K_{{\cal H}}(\vec{x},\vec{y};\beta )=0 \label{eq:heateq}
\end{equation}
with initial condition
\begin{equation}
K_{{\cal H}}(\vec{x},\vec{y};0)={\mathbb I}\cdot
\delta^{(2)}(\vec{x}-\vec{y})  \quad .\label{eq:inicond}
\end{equation}
Because
\[
K_{{\cal H}_0}(\vec{x},\vec{y};\beta)={e^{-\beta}\over
4\pi\beta}\cdot{\mathbb I}\cdot e^{-\frac{|\vec{x}-\vec{y}|
}{4\beta}}
\]
is the heat kernel for the Klein-Gordon operator ${\cal H}_0$, it
is convenient to write the heat kernel for ${\cal H}$ in the form:
\begin{equation}
K_{{\cal H}}(\vec{x},\vec{y};\beta)=C_{\cal
H}(\vec{x},\vec{y};\beta)K_{{\cal H}_0}(\vec{x},\vec{y};\beta)
\label{eq:tru1}
\end{equation}
with $C_{\cal H}(\vec{x},\vec{x};0)={\mathbb I}$ \cite{Roe}.
Substituting (\ref{eq:tru1}) into (\ref{eq:heateq}) we find that
$C_{\cal H}(\vec{x},\vec{y};\beta)$ solves the transfer equations:
\begin{eqnarray}
\left\{ {\partial\over\partial\beta}{\mathbb
I}+{x_k-y_k\over\beta}(\partial_k{\mathbb I}-{1\over
2}Q_k)-\bigtriangleup{\mathbb I}+  \right.&& \nonumber \\
\left. +Q_k\partial_k+V  \rule[-0.1in]{0.in}{0.2in}
\right\}C_{\cal H}(\vec{x},\vec{y};\beta)=0 && \quad
.\label{eq:tran}
\end{eqnarray}
The high-temperature expansion
\[
C_{\cal H}(\vec{x},\vec{y};\beta)=\sum_{n=0}^\infty
c_n(\vec{x},\vec{y};{\cal H})\beta^n
\]
trades the PDE (\ref{eq:tran}) by the recurrence relations
\begin{eqnarray}
[n{\mathbb I}+(x_k-y_k)(\partial_k{\mathbb I}-{1\over
2}Q_k)]c_n(\vec{x},\vec{y};{\cal H}) =&& \nonumber
\\=[\bigtriangleup{\mathbb I}
-Q_k\partial_k-V]c_{n-1}(\vec{x},\vec{y};{\cal H}) &&
\label{eq:rec}
\end{eqnarray}
among the local coefficients with $n\geq 1$, with the initial
condition $c_0(\vec{x},\vec{x};{\cal H})={\mathbb I}$. Taking into
account that
\begin{eqnarray}
{\rm Tr}e^{-\beta{\cal H}}&=&{e^{-\beta}\over
4\pi\beta}\sum_{n=0}^\infty\sum_{a=1}^4\int \, d^2x
\,[c_n]_{aa}(\vec{x},\vec{x};{\cal H})\beta^n  = \nonumber \\
&=& {e^{-\beta}\over 4\pi\beta} \sum_{n=0}^\infty\beta^n c_n({\cal
H})\qquad ,
\end{eqnarray}
where we have defined the Seeley coefficients as
\[
c_n({\cal H})=\sum_{a=1}^4 \int \, d^2x
\,[c_n]_{aa}(\vec{x},\vec{x};{\cal H}) \qquad ,
\]
and that the first local coefficient can be easily computed
\[
c_1(\vec{x},\vec{x};{\cal H})=-V(\vec{x}) \qquad ,
\]
by applying these formulas to the ${\cal H}^\pm$ operators we
obtain in the $\beta=0$ -infinite temperature- limit :
\begin{eqnarray*}
{\rm ind}{\cal D}&=&{1\over 4\pi}\left\{c_1({\cal H}^+)-c_1({\cal
H}^-)\right\}=\\&=&{1\over \pi}\int d^2x \left(\frac{\partial
V_2}{\partial x_1}-\frac{\partial V_1}{\partial x_2}\right)
(\vec{x})=2l,
\end{eqnarray*}
i.e., the dimension of the self-dual vortex moduli space is $2l$.
Physically, this means that there are solutions, if $\kappa=1$,
for any location of the $l$-vortex centers in the plane
\cite{jatb}; all static configurations of self-dual $l$-vortices
can thus be interpreted as states of neutral equilibrium.

\subsection{Seeley coefficients}

Computation of the coefficients of the asymptotic expansion is a
difficult task; to start with, the order two local coefficient
reads:
\begin{eqnarray*}
&& c_2(\vec{x},\vec{x};{\cal H})= -{1\over 6}\bigtriangleup
V(\vec{x})+{1\over
12}Q_k(\vec{x})Q_k(\vec{x})V(\vec{x})-\\
&&- {1\over 6}\partial_kQ_k(\vec{x})V(\vec{x})+{1\over
6}Q_k(\vec{x})\partial_kV(\vec{x})+{1\over 2}V^2(\vec{x})\quad .
\end{eqnarray*}
Complexity increases strongly for high-order local coefficients .

The recurrence relation (\ref{eq:rec}) allows us to express
$c_n(\vec{x},\vec{y},{\cal H})$ and its derivatives in terms of
all the $c_k(\vec{x},\vec{y};{\cal H})$ with $k\leq n$ and their
derivatives. One passes from this information to the values of the
Seeley coefficients $c_n({\cal H})$ in two steps. First, one must
reach the subtle $\vec{y}\rightarrow \vec{x}$ limit. In this
analytical manoeuvre the partial derivatives of
$c_n(\vec{x},\vec{y};{\cal H})$ at $\vec{y}=\vec{x}$
\[
{}^{(\alpha_1,\alpha_2)}C_n^{ab}(\vec{x})=\lim_{\vec{y}\rightarrow
\vec{x}} \frac{\partial^{\alpha_1+\alpha_2}
[c_n]_{ab}(\vec{x},\vec{y};{\cal H})}{\partial
x_1^{\alpha_1}\partial x_2^{\alpha_2}}
\]
play a prominent r$\hat{{\rm o}}$le. Note also that:
\[
[{c}_n]_{ab}(\vec{x},\vec{x};{\cal
H})={}^{(0,0)}C_n^{ab}(\vec{x})\qquad .
\]
In the $\vec{y}\rightarrow\vec{x}$ limit the recurrence relation
(\ref{eq:rec}) becomes :
\begin{widetext}
\begin{eqnarray} (k+\alpha_1&+&\alpha_2+1)
{}^{(\alpha_1,\alpha_2)}C_{k+1}^{ab}(\vec{x})=
{}^{(\alpha_1+2,\alpha_2)}C_{k}^{ab}(\vec{x})+
{}^{(\alpha_1,\alpha_2+2)}C_{k}^{ab}(\vec{x})- \nonumber \\
&&-\sum_{d=1}^N \sum_{r=0}^{\alpha_1}\sum_{t=0}^{\alpha_2}
{\alpha_1 \choose r} {\alpha_2 \choose t} \left[
\frac{\partial^{r+t} Q^{ad}_1}{\partial x_1^r\partial x_2^t}
{}^{(\alpha_1-r+1,\alpha_2-t)}C_{k}^{db}(\vec{x})+\frac{\partial^{r+t}
Q^{ad}_2}{\partial x_1^r\partial x_2^t}
{}^{(\alpha_1-r,\alpha_2-t+1)}C_{k}^{db}(\vec{x})
\right]+ \nonumber \\
&&+\frac{1}{2}\sum_{d=1}^N
\sum_{r=0}^{\alpha_1-1}\sum_{t=0}^{\alpha_2} \alpha_1{\alpha_1-1
\choose r} {\alpha_2 \choose t}  \frac{\partial^{r+t}
Q^{ad}_1}{\partial x_1^r\partial x_2^t}
{}^{(\alpha_1-1-r,\alpha_2-t)}C_{k+1}^{db}(\vec{x})+ \label{eq:recur}\\
&&+\frac{1}{2}\sum_{d=1}^N
\sum_{r=0}^{\alpha_2-1}\sum_{t=0}^{\alpha_1} \alpha_2{\alpha_2-1
\choose r} {\alpha_1 \choose t}  \frac{\partial^{r+t}
Q^{ad}_2}{\partial x_1^t\partial x_2^r}
{}^{(\alpha_1-t,\alpha_2-1-r)}C_{k+1}^{db}(\vec{x})- \nonumber
\\&&-\sum_{d=1}^N
\sum_{r=0}^{\alpha_2}\sum_{t=0}^{\alpha_1}{\alpha_1 \choose
t}{\alpha_2\choose r}
 \frac{\partial^{r+t}
V^{ad}}{\partial x_1^t\partial x_2^r}
{}^{(\alpha_1-t,\alpha_2-r)}C_k^{db}(\vec{x}) \nonumber \qquad .
\end{eqnarray}
\end{widetext}
The initial condition $c_0(\vec{x},\vec{x};{\cal H})={\mathbb I}$
means that all the ${}^{(\beta,\gamma)}C_0^{ab}(\vec{x})$ vanish
except ${}^{(0,0)}C_0^{aa}(\vec{x})=1$ for $a=1,2,\cdots,N$.
Starting from these conditions one computes all the
${}^{(\beta,\gamma)}C_n^{ab}(\vec{x})$ local coefficients by using
(\ref{eq:recur}). For instance, in order to obtain
${}^{(0,0)}C_6^{ab}(\vec{x})$ for ${\cal H}^+$ we need
${}^{(\beta,\gamma)}C_5^{ab}(\vec{x})$ for $\beta,\gamma=0,1,2$ as
data, which in turn can be calculated from
${}^{(\beta,\gamma)}C_4^{ab}(\vec{x})$ for
$\beta,\gamma=0,1,2,3,4$, and so forth. Evaluation of
${}^{(0,0)}C_6^{ab}(\vec{x})$ requires knowledge of 4032 local
coefficients !!!. In general, the rule is: knowledge of
${}^{(0,0)}C_n^{ab}(\vec{x})$ amounts to knowledge of
$\frac{8}{3}(n+1)(n+2)(4n+3)$
${}^{(\beta,\gamma)}C_k^{ab}(\vec{x})$ local coefficients with
$k\leq n$. The second step is much simpler: simple numerical
integration of $\sum_{a=1}^4{}^{(0,0)}C_n^{aa}(\vec{x})$ over the
plane.

\section{Quantum oscillations of self-dual vortices}

Standard lore in the semi-classical quantization of solitons tells
us that the one-loop mass shift comes from the Casimir energy plus
the contribution of the mass renormalization counter-terms:
$\Delta M_V=\Delta M_V^C+\Delta M_V^R$.

\subsection{Casimir energy and vortex mass renormalization
counter-terms}

By expanding the static energy (\ref{eq:he}) of the AHM around
self-dual vortex solutions one obtains, up to second-order in
$\xi$ in the background gauge:
\begin{eqnarray*}
E&+&{v^2\over 2}\int \, d^2\vec{x} \, \left[\partial_j
a_j-\psi_1\varphi_2+\psi_2\varphi_1\right]^2 \\&\simeq& \pi |l|
v^2+{1\over 2}\int \, d^2\vec{x} \, \xi^T{\cal H}^+ \xi +{\cal
O}(\xi^3) \qquad .
\end{eqnarray*}
Also, the ghosts -arising when the quantization procedure is
performed in the background gauge- contribute negatively to the
energy:
\[
E^{{\rm Ghost}}={v^2\over 2}\int \, d^2\vec{x} \,
\left[\chi^*\left(-\bigtriangleup+|\psi|^2\right)\chi+\psi^*\varphi\chi^*\chi\right]\quad
.
\]
Thus, the vortex Casimir energy is the sum of the Casimir energies
of the bosonic $a_1,a_2,\varphi_1,\varphi_2$ fluctuations around
the vortex minus the Casimir energy of the fermionic fluctuation
$\chi$; the ordinary -non-matrix- Schrodinger operator ruling the
ghost fluctuation around the vortex is:
\[
{\cal H}^G=-\bigtriangleup+|\psi|^2 \qquad .
\]
$\varphi_2$ is a pure gauge oscillation but its contribution is
killed by the negative ghost contribution. The same applies
 for
the vacuum Casimir energy: the Goldstone boson Casimir energy is
canceled by the ghost Casimir energy, the trace of the square root
of ${\cal H}_0^G=-\bigtriangleup +1$. In sum, the vortex Casimir
energy measured with respect to the vacuum Casimir energy is given
by the formal formula:
\begin{eqnarray*}
\Delta M_V^C&=&{\hbar m\over 2}\left[{\rm STr}^*\left({\cal
H}^+\right)^{{1\over 2}}-{\rm STr}\left({\cal H}_0\right)^{{1\over
2}}\right] \\ {\rm STr}^*\left({\cal H}^+\right)^{{1\over
2}}&=&\hspace{0.3cm}{\rm Tr}^* \left({\cal H}^+\right)^{{1\over
2}}-{\rm Tr}\left({\cal H}^G\right)^{{1\over 2}}\\ {\rm
STr}\left({\cal H}_0\right)^{{1\over 2}}&=&\hspace{0.3cm}{\rm
Tr}\left({\cal H}_0\right)^{{1\over 2}}-{\rm Tr}\left({\cal H}^G_0
\right) \qquad .
\end{eqnarray*}
The star means that the $2l$ zero eigenvalues of ${\cal H}^+$ must
be subtracted because zero modes only enter at two-loop order.

In the minimal subtraction renormalization scheme, finite
renormalizations are adjusted in such a way that the critical
point $\kappa^2=1$ is reached at first-order in the loop
expansion. Therefore, (\ref{eq:ct1}) and (\ref{eq:ct2}) tell us
that the contribution of the mass renormalization counter-terms to
the vortex mass is:
\begin{eqnarray*}
\Delta M_V^R&=&\Delta M_{c.t.}^S+\Delta M_{c.t.}^A=\hbar\, m\,I(1)
\, \Sigma (\psi,V_k)\\
\Sigma (\psi,V_k)&=&\int \, dx^2 \, [(1-|\psi|^2)-{1\over 2}
V_kV_k]\hspace{0.2cm},
\end{eqnarray*}
and the divergent integral $I(1)$ can be written in the form
\[
I(1)={1 \over 2}\int {d^2 {\vec k}\over (2 \pi)^2} {1\over
\sqrt{\vec k \cdot \vec k +1}}
\]
after applying the residue theorem to integration in the complex
$k_0$-plane.

\subsection{Zeta function regularization of Casimir energies and
self-energy graphs}

We regularize both infinite quantities $\Delta M_V^C$ and $\Delta
M_V^R$ by means of generalized zeta functions. From the spectral
resolution of a Fredholm operator ${\cal H}$
\[
{\cal H}\xi_n=\lambda_n\xi_n \qquad ,
\]
one defines the generalized zeta function as the series
\[
\zeta_{\cal H}(s)=\sum_n {1\over\lambda_n^s} \qquad ,
\]
which is a meromorphic function of the complex variable $s$
\cite{Gilkey}, \cite{Roe}. We can then hope that, despite their
continuous spectra, our operators fit in this scheme, and write:
\begin{eqnarray*}
\Delta M_V^C (s)&=&\frac{\hbar\mu}{2}\left({\mu^2\over
m^2}\right)^s\left\{\left(\zeta_{{\cal H}^+}(s)-\zeta_{{\cal
H}^G}(s)\right)+ \right. \\ &&+\left. \left(\zeta_{{\cal
H}_0^G}(s)-\zeta_{{\cal
H}_0}(s)\right)\right\} \\
\Delta M_V^R(s)& = &{\hbar\over m L^2} \zeta_{{\cal H}_0} (s)
\Sigma (\psi,V_k)
\end{eqnarray*}
where
\[
\zeta_{{\cal H}_0} (s) = {m^2 L^2 \over 4 \pi} {\Gamma (s-1) \over
\Gamma(s)}
\]
and $\mu$ is a parameter of inverse length dimensions. Note that
\[
\Delta M_V^C=\lim_{s\rightarrow -\frac{1}{2}}\Delta M_V^C(s)\quad
, \quad \Delta M_V^R=\lim_{s\rightarrow \frac{1}{2}}\Delta
M_V^R(s)
\]
and
\[
I(1)=\lim_{s\rightarrow{1\over 2}}{1\over 2m^2L^2}\zeta_{{\cal
H}_0}(s)
\]
on a square of area $L^2$.

Together with the high-temperature expansion, the Mellin transform
of the heat trace
\[
\zeta_{{\cal H}}(s)={1\over\Gamma(s)}\int_0^\infty \, d\beta \,
\beta^{s-1} \, {\rm Tr} \, e^{-\beta {\cal H}}
\]
shows that
\[
\zeta_{{\cal H}}(s)={1\over\Gamma(s)}\sum_{n=0}^\infty \int_0^1 \,
d\beta \, \beta^{s+n-2}c_n({\cal
H})e^{-\beta}+{1\over\Gamma(s)}B_{{\cal H}}(s)
\]
is the sum of meromorphic and entire --$B_{\cal H}(s)$-- functions
of $s$. Neglecting the entire parts and keeping a finite number of
terms, $N_0$, in the asymptotic series for $\zeta_{\cal H}(s)$, we
find the following approximations for the generalized zeta
functions concerning the differential operators ${\cal H}^+$ and
${\cal H}^G$ relevant to our problem:
\begin{eqnarray*}
\zeta_{{\cal H}^+} (s) - \zeta_{{\cal H}_0} (s) & \simeq &
\sum_{n=1}^{N_0} c_n({\cal H}^+) \cdot
{\gamma[s+n-1,1] \over 4 \pi \Gamma(s)} \\
 \zeta_{{\cal H}_0^G} (s) -\zeta_{{\cal H}^G} (s)
& \simeq & - \sum_{n=1}^{N_0} c_n ({\cal H}^G) \cdot
{\gamma[s+n-1,1] \over 4 \pi \Gamma(s)} \quad ;
\end{eqnarray*}
$\gamma[s+n-1,1]=\int_0^1 \, d\beta \, \beta^{s+n-2}e^{-\beta} $
is the incomplete gamma function, with a very well known
meromorphic structure.

Regarding one-dimensional kinks, see \cite{Aai1}, \cite{Aai2},
\cite{Aai3}, the contributions of $c_0 ({\cal H}^+)$ and
$c_0({\cal H}^G)$ to $\zeta_{{\cal H}^+}(s)$ and $\zeta_{{\cal
H}^G}(s)$ are respectively canceled by $\zeta_{{\cal H}_0}(s)$ and
$\zeta_{{\cal H}^G_0}(s)$; i.e., renormalization of zero point
vacuum energies takes care of the $c_0({\cal H}^+)$ and $c_0({\cal
H}^G)$ contributions to the vortex Casimir energy. Note, however,
that, in contrast to the (1+1)-dimensional case, the value
$s=-{1\over 2}$ for which we shall obtain the Casimir energy is
not a pole. To compute the vortex Casimir energy one can first
take the $s=-{1\over 2}$ limit and then subtract the vacuum
Casimir energy regularized by this procedure; a finite answer for
the kink Casimir energy is only reached if one first subtracts the
vacuum Casimir energy of the one-dimensional system.

\subsection{One-loop mass shift formula}

Writing as $\bar{c}_n=c_n({\cal H}^+)-c_n ({\cal H}^G)$ the
difference between the Seeley coefficients of ${\cal H}^+$ and
${\cal H}^G$ for vorticity $l$, we check that the contribution of
the first coefficient to the Casimir energy
\[
\Delta M_V^{(1)C} (s)   \simeq {\hbar \over 2} \mu \left( {\mu^2
\over m^2}\right)^s \bar{c}_1 \cdot {\gamma[s,1/2] \over 4 \pi
\Gamma(s)}
\]
is finite at the $s\rightarrow -\frac{1}{2}$ limit
\[
\Delta M_V^{(1)C} (-1/2)  \simeq - {\hbar m \over 4 \pi } \Sigma
(\psi,V_k) \cdot {\gamma[-1/2,1] \over \Gamma(1/2)}
\]
and exactly cancels the contribution of the mass renormalization
counter-terms --also finite for $s={1\over 2}$--:
\begin{eqnarray*}
\Delta M_V^R (s)& \simeq &{\hbar m \over 4 \pi} \cdot \Sigma
(\psi,V_k) \cdot {\gamma[s-1,1]\over  \Gamma(s)} \\
\Delta M_V^R (1/2) & \simeq & {\hbar m \over 4 \pi} \cdot \Sigma
(\psi,V_k) \cdot
 {\gamma[-1/2,1]\over \Gamma(1/2)} \, .
\end{eqnarray*}
Our choice of a minimal subtraction scheme not only arranges
finite renormalizations in such a way that self-duality holds for
$\kappa=1$ at the one-loop order, but also fits in with the
criterion that the mass renormalization counter-terms must kill
the contribution to the Casimir energy of the first Seeley
coefficients for the heat trace expansions of the operators ${\cal
H}^+$, ${\cal H}^G$, ${\cal H}_0$, ${\cal H}_0^G$. The same
cancellation happens for kinks only if the mode number cut-off
regularization procedure, see \cite{Reb1}, \cite{Aai1} and
\cite{vanN}, is applied.

Subtracting the contribution of the $2l$ zero modes,
\begin{eqnarray*}
 \Delta M_{V}& =&{\hbar m\over 2}\lim_{s\rightarrow
-\frac{1}{2}}\left[-\frac{2l}{\Gamma(s)}\int_0^1 d\beta
\beta^{s-1}+ \right.  \\ && \left. + \sum_{n=2}^{N_0} \bar{c}_n
{\gamma[s+n-1,1] \over 4 \pi \Gamma(s)}\right]
\end{eqnarray*}
we finally obtain the following formula for the vortex mass shift:
\begin{equation}
\Delta M_V= -{\hbar m \over 2} \left[ \frac{1}{8\pi\sqrt{\pi}}
\sum_{n=2}^{N_0} \bar{c}_n \gamma[n-\frac{3}{2},1]+\frac{2
l}{\sqrt{\pi}} \right] \, . \label{eq:vorm}
\end{equation}

\section{One-loop mass shifts}

\subsection{Local coefficients for cylindrically symmetric vortices}

We shall apply these formulae to cylindrically symmetric vortices.
The heat kernel local coefficients, however, depend on successive
derivatives of the solution. This dependence can increase the
error in the estimation of these local coefficients because we
handle an interpolating polynomial as the numerically generated
solution, and the successive derivations with respect to $r$ of
such a polynomial introduces inaccuracies. Indeed this operation
is plugged into the algorithm that generates the local
coefficients in order to speed up this process. It is thus of
crucial importance to use the first-order differential equations
(\ref{eq:rrfo}) in order to eliminate the derivatives of the
solution and write the local coefficients as expressions depending
only on the fields. We find:
\begin{eqnarray*}
\frac{\partial\psi_1}{\partial x_1}&=&\frac{l
f(r)}{r}\left[\cos\theta \cos l\theta(1-\alpha(r))+\sin\theta \sin l\theta \right]\\
\frac{\partial\psi_1}{\partial x_2}&=&\frac{l
f(r)}{r}\left[\sin\theta \cos l\theta(1-\alpha(r))- \cos\theta \sin l\theta \right]\\
\frac{\partial\psi_2}{\partial x_1}&=&\frac{l
f(r)}{r}\left[\cos\theta \sin l\theta(1-\alpha(r))- \sin\theta \cos l\theta \right]\\
\frac{\partial\psi_2}{\partial x_2}&=&\frac{l
f(r)}{r}\left[\sin\theta \sin l\theta(1-\alpha(r))+ \cos\theta
\cos l\theta \right] \\ \frac{\partial V_1}{\partial
x_1}&=&\sin\theta \cos
\theta\left[\frac{2l f(r)\alpha(r)}{r}+{1\over 2}(f^2(r)-1)\right]\\
\frac{\partial V_1}{\partial x_2}&=&-l \cos
2\theta\frac{\alpha(r)}{r^2}+{1\over 2}\sin^2\theta(f^2(r)-1)\\
\frac{\partial V_2}{\partial x_1}&=&-l \cos
2\theta\frac{\alpha(r)}{r^2}-{1\over 2} \cos^2\theta(f^2(r)-1)\\
\frac{\partial V_2}{\partial x_2}&=&-\sin\theta \cos
\theta\left[\frac{2l f(r)\alpha(r)}{r}+{1\over 2}(f^2(r)-1)\right]
\end{eqnarray*}
for self-dual ANO vortices with generic (positive) vorticity $l$.

The recurrence formula now gives the local coefficients of the
asymptotic expansion in terms of $f(r)$ and $\alpha(r)$, e.g.,
\begin{widetext}
{\footnotesize\begin{eqnarray*} {\rm tr}
[c_1](\vec{x},\vec{x};{\cal H^+})&=& 5[1-f^2(r)]-\frac{2}{r^2} l^2
\alpha^2
(r) \\
{\rm tr}[c_2](\vec{x},\vec{x};{\cal H^+})&=&
\frac{1}{12r^4}\left\{ 37 r^4 +4 l^4 \alpha^4
(r)+8(7l^2r^2-8r^4)f^2(r)+27r^4 f^4(r)- \right.
\\ && \left. -8lr^2 \alpha(r)[-1+(1+13
l)f^2(r)]+8l^2\alpha^2(r)(-2-3r^2+9r^2f^2(r)) \right\} \\
{\rm tr}[c_3](\vec{x},\vec{x};{\cal H^+})&=& \frac{1}{120 r^6}
\left\{ -4l^6 \alpha^6(r)-4 l^3 r^2 \alpha^3(r) [14 +(-132+167
l)f^2(r)]+4l^4 \alpha^4(r)(20+9r^2+32r^2f^2(r))-2lr^2 \alpha(r)
[-4(16+9r^2)+ \right.
\\ && + (64+96 l-472 l^2+344 l^3+88l^2+243 l r^2)
f^2(r)+(-52+109l)r^2 f^4(r)]+ l^2 \alpha^2(r) [-256-144
r^2-117r^4+ \\ && + 2r^2(88-548 l+516l^2+183r^2)f^2(r)  + 99r^4
f^4(r)]+r^2[r^2(-16+151r^2)+(-320l^3+160l^4+32r^2+48lr^2-
\\ && \left.-321 r^4 +
8l^2(20+39r^2))f^2(r)+r^2(-16-48
l+44l^2+199r^2)f^4(r)-29r^4f^6(r)] \right\} \qquad .
\end{eqnarray*}}
\end{widetext}
We have explicitly given only the first three local coefficients
of the heat kernel expansion for ${\cal H}^+$ because the
complexity of the expressions increases with $n$ enormously.
Additionally, {\footnotesize\begin{eqnarray*}
&& c_1(\vec{x},\vec{x};{\cal H}^G)= 1-f^2(r)\\
c_2(\vec{x},\vec{x};{\cal H}^G)&=&\frac{-1}{6r^2}\left\{ [4 l^2+ 5
r^2-8 l^2
\alpha(r) +4 l^2 \alpha^2(r)]f^2(r)+ \right. \\
&& \left. \hspace{0.6cm} -3 r^2-2r^2f^4(r) \right\} \\
c_3(\vec{x},\vec{x};{\cal H}^G)&=& \frac{1}{60 r^4} \left\{ 10
r^4-[-32 l^3+16 l^4+8lr^2+23 r^4+ \right. \\  + 16 l^2 (1+r^2)&-&
8l(-12l^2+8l^3+r^2+4l(1+r^2))\alpha(r)+ \\  + 16
l^2(1-6l+6l^2&+&r^2)\alpha^2(r)+ 32(1-2l)l^3\alpha^3(r)+ \\ +
16l^4\alpha^4(r)]f^2(r)&+& r^2[8l+16l^2+17r^2+16l^2\alpha^2(r)-
\\ &&-8l(1+4l)\alpha(r)\left. ] f^4(r)-4r^4f^6(r)
\right\}
\end{eqnarray*}}

\noindent are the first three local coefficients for the heat
kernel expansion for the ghost operator ${\cal H}^G$.

Plugging these expressions into the partially analytical partially
numerical solution for $f(r)$ and $\alpha(r)$, it is possible to
compute the local coefficients and integrate them numerically over
the whole plane.

\subsection{Mass shift for vorticities $l=1$, $l=2$, $l=3$, $l=4$}

Finally, the one-loop quantum correction of the vortex solution
with vorticity $l$ is given by formula (\ref{eq:vorm})
\[
\Delta M_V= -{\hbar m \over 2} \left[ \frac{1}{8\pi\sqrt{\pi}}
\sum_{n=2}^{N_0} \bar{c}_n \gamma[n-\frac{3}{2},1]+\frac{2
l}{\sqrt{\pi}} \right]
\]
Using the Mathematica environment in a modest PC we have obtained
the coefficients shown in Tables IV and V,

\begin{center}
\begin{table}[h]
  \caption{Seeley Coefficients for $l=1,2$.}
  \label{table:tabla1}
\begin{tabular}{|c|cc|cc|}
\hline & \multicolumn{2}{|c|}{$l=1$} & \multicolumn{2}{|c|}{$l=2$}
\\ \hline
$n$ & $ {c}_n({\cal H}^+)$ & $ {c}_n({\cal H}^G)$ & $ {c}_n({\cal
H}^+)$ & $ {c}_n({\cal H}^G)$
\\ \hline
2 & 30.36316 & 2.60773 & 61.06679 & 6.81760    \\
3 & 12.94926 & 0.31851 & 25.61572 & 1.34209    \\
4 &  4.22814 &  0.022887 & 8.21053 & 0.20481   \\
5 & 1.05116 &  0.0011928 & 2.02107 & 0.023714  \\
6 &  0.20094 & 0.00008803 & 0.40233 & 0.002212  \\
\hline
\end{tabular}
\end{table}
\end{center}

\begin{center}
\begin{table}[h]
  \caption{Seeley Coefficients for $l=3,4$.}
  \label{table:tabla1}
\begin{tabular}{|c|cc|cc|}
\hline & \multicolumn{2}{|c|}{$l=3$} & \multicolumn{2}{|c|}{$l=4$}
\\ \hline
$n$ & $ {c}_n({\cal H}^+)$ & $ {c}_n({\cal H}^G)$ & ${c}_n({\cal
H}^+)$ & $ {c}_n({\cal H}^G)$
\\ \hline
2 & 90.20440 & 11.51035 & 118.67540 & 16.46895 \\
3 & 36.68235 &  2.60898 &  46.01141 & 4.00762 \\
4 & 11.69979 &  0.46721 &  14.64761 & 0.77193 \\
5 & 2.86756  & 0.067279 &   3.58906 & 0.11747 \\
6 & 0.566227  & 0.0079269 & 0.667202 & 0.01620 \\
\hline
\end{tabular}
\end{table}
\end{center}

We remark that formula (\ref{eq:vorm}) depends on the number $N_0$
chosen to cut the asymptotic heat kernel expansions. We have no
means of determining the optimum value for $N_0$, but in practice
we can only cope with a small $N_0$ value; a big $N_0$ would
require the computation of an enormous number of local
coefficients. Nevertheless, the choice $N_0=6$ is acceptable. The
behavior of the asymptotic series in (\ref{eq:vorm}) is given in
Table VI:
\begin{table}[h]
  \caption{Convergence of the asymptotic series in units of $\hbar m$.}
  \label{table:tabla1}
\begin{tabular}{|c|cccc|}  \hline
$N_0$ & $\Delta M_V (N_0)$ & $\Delta M_V (N_0)$& $\Delta M_V
(N_0)$ & $\Delta M_V (N_0)$ \\
 & $l=1$ & $l=2$ & $l=3$ & $l=4$
\\ \hline
2 & -1.02951 & -2.03787  &  -3.01187 & -3.97025 \\
3 & -1.08323 & -2.14111  &  -3.15680 & -4.14891 \\
4 & -1.09270 & -2.15913  &  -3.18208 & -4.18014 \\
5 & -1.09427 & -2.16212  &  -3.18628 & -4.18534 \\
6 & -1.09449 & -2.16257  &  -3.18690 & -4.18606 \\ \hline
\end{tabular}
\end{table}

The convergence up to the sixth order in the asymptotic expansion
is very good. In the case of $\lambda (\phi)^4_2$ kinks we found
agreement between the result obtained by this method and the exact
result up to the fourth decimal figure, see \cite{Aai1}, by
choosing $N_0=10$.

There are reasons to expect this behavior on general analytical
grounds. Truncation of the asymptotic expansion of the heat
function at order $N_0$ produces an error of order $\beta^{N_0}$,
which in turn leads to an error proportional to
$\gamma[N_0-{1\over 2},1]\simeq {1\over N_0-{1\over 2}}$, for
$N_0$ large, in the computation of the $\zeta_{{\cal
H}^+}(-{1\over 2})$ zeta function, see \cite{Gilkey} Section 1.10.
In fact, the rate of convergence is improved in our problem by the
the smallness of the $c_n$ coefficients, see Tables IV and V, for
large n. This smallness is due to the fact that, when $n$
increases, higher and higher powers of partial derivatives of the
field profiles of increasing order enter in the computation of
$c_n$. The vortex solutions, however, are as regular and smooth as
allowed by the topology. Therefore, the admitted error by cutting
the mass shift formula at $N_0=6$ is especially small for low
vorticities.

In Table VII we give the one-loop quantum corrections for the
vortex solutions up to $l=4$, whereas we plot the correction in
the figure as a function of the magnetic flux. The broken line
(linear function) represents the hypothetical situation in which
each magnetic flux quantum would contribute with the same
correction. Hence, this is almost - within the error margin- the
situation that we have found.
\begin{table}
\caption{One-Loop Quantum Mass Correction to the vortex with
vorticity $l=1,2,3,4$.}
\begin{center}
\begin{tabular}{|c|c|}
\hline $l$ & $\Delta M_V/\hbar m$ \\ \hline 1 & -1.09449  \\
2 & -2.16257 \\ 3 & -3.18690 \\  4 & -4.18606 \\
\hline
\end{tabular} \hspace{0.7cm}
\begin{tabular}{c}\includegraphics[height=2.2cm]{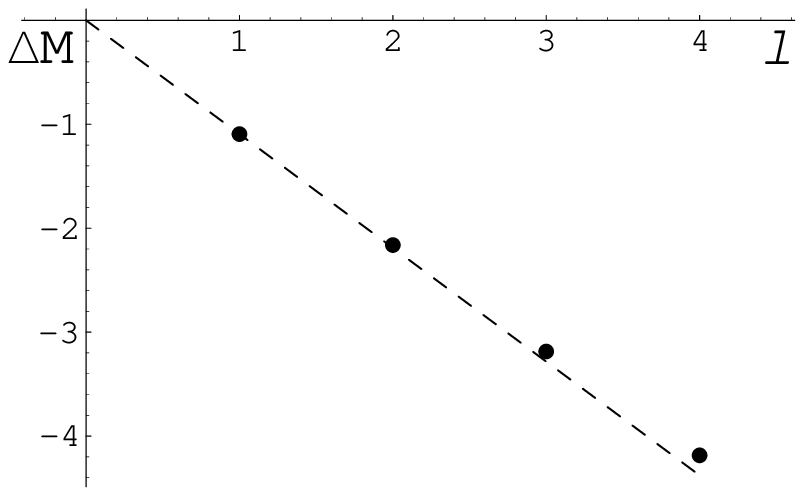}
\end{tabular}
\end{center}
\end{table}

These results, however, do not allow us to answer the question of
whether or not the classical degeneracy with respect to the vortex
centers observed at the classical level also holds at one-loop
order. The figure in Table VII seems to suggest that the mass
shift of $l$ well separated vortices is equal -modulo errors- to
$l$ times the mass shift of a single vortex, but we do not know in
what direction the errors run.

\subsection{Mass shift for solutions with two separate vortices}

 We now offer two Tables, VIII and IX, where Seeley
coefficients and the quantum corrections are given for two-vortex
solutions with intermediate separations $d=1$, $d=2$, and $d=3$
between superimposed vortices, $\omega=0$ in
(\ref{eq:fun1})-(\ref{eq:fun2}), and well separated vortices,
$\omega=1$ in (\ref{eq:fun1})-(\ref{eq:fun2}). The coefficients of
the asymptotic expansion are computed only up to third order
because much more computation time is required. Also, we stress
that in this situation, with no cylindrical symmetry, we expect
not so good results because there are two more important sources
of errors: first, the variational solutions with two separate
vortices are far less exact than the solution with $l=2$ and
cylindrical symmetry. Second, even though another numerical method
would be used in the search of vortex solution we would run in
difficulties; there is no way to avoid the use of partial
derivatives in the calculation of the coefficients because the
vortex equations alone are not enough.
\begin{center}
\begin{table}[h]
  \caption{Seeley Coefficients for $d=1,2,3$.}
  \label{table:tabla11}
\begin{tabular}{|c|cc|cc|cc|}
\hline & \multicolumn{2}{|c|}{$d=1$} & \multicolumn{2}{|c|}{$d=2$}
&  \multicolumn{2}{|c|}{$d=3$}
\\ \hline
$n$ & $ {c}_n({\cal H}^+)$ & $ {c}_n({\cal H}^G)$ & $ {c}_n({\cal
H}^+)$ & $ {c}_n({\cal H}^G)$ & $ {c}_n({\cal H}^+)$ & $
{c}_n({\cal H}^G)$
\\ \hline
2 & 61.0518 & 6.81277 & 58.3359 & 6.46609  & 57.3420 & 6.03872  \\
3 & 25.6137 & 1.33822 & 24.5050 & 1.23466  & 24.1187 & 1.02031  \\
\hline
\end{tabular}
\end{table}
\end{center}

\begin{table}[h]
  \caption{Convergence of the asymptotic series.}
  \label{table:tabla12}
\begin{tabular}{|c|ccc|}  \hline
$N_0$ & $\Delta M_V (N_0)/\hbar m$ & $\Delta M_V (N_0)/\hbar m$&
$\Delta M_V
(N_0)/\hbar m$ \\
 & $d=1$ & $d=2$ & $d=3$
\\ \hline
2 & -2.03770 & -1.99798  &  -1.98848  \\
3 & -2.14095 & -2.09695  &  -2.08672  \\
\hline
\end{tabular}
\end{table}

\section{Summary and outlook}

The one-loop mass shifts of superimposed vortices with low
magnetic fluxes are:
\begin{eqnarray*}
M_V^{l=1}&=&m\left(\frac{\pi v}{e}-1.09427
\hbar\right)+o(\hbar^2)\\
M_V^{l=2}&=&2 m\left(\frac{\pi v}{e}-1.08106
\hbar\right)+o(\hbar^2)\\
M_V^{l=3}&=&3 m\left(\frac{\pi v}{e}-1.06230
\hbar\right)+o(\hbar^2) \\
M_V^{l=4}&=& 4 m\left(\frac{\pi
v}{e}-1.04651\hbar\right)+o(\hbar^2).
\end{eqnarray*}
Much less precise results are also provided for two-vortices with
separate centers. This is to be compared with the supersymmetric
result:
\[
{\cal M}_V^{l}= |l| m \left(\frac{\pi v}{e}-0.5000
\hbar\right)+o(\hbar^2) \qquad ,
\]
see \cite{Vass} and \cite{Reb}. We notice that the one-loop
correction due to bosonic fluctuations of self-dual vortices is
almost twice the correction arising in the supersymmetric system
coming only from mass renormalization counterterms when proper
SUSY-preserving boundary conditions are imposed. The same
proportion holds between one-loop corrections to sine-Gordon and
$\phi^4$ kink masses in the non-supersymmetric and supersymmetric
frameworks, see \cite{Schf} and \cite{Wimm}.

It seems plausible that a similar method can successfully be
applied to compute the one-loop mass shift for self-dual
Chern-Simons-Higgs vortices, see \cite{Wein1}-\cite{GM99}. A
Hamiltonian formalism in the topological sectors of the
first-order CSH Lagrangian system should be first developed. More
ambitious, generalized zeta functions of $12\times 12$ matrix PDE
operators in three variables are essential in computing the
one-loop mass shift to BPS monopoles. Thus, our procedure opens a
door to calculate quantum corrections to BPS monopole masses in a
${\cal N}=0$ bosonic setting to be contrasted with the ${\cal
N}=2$ and ${\cal N}=4$ supersymmetric results of \cite{Reb2} and
\cite{Reb3}. \medskip

\section*{Acknowledgements}Partially financed by the Spanish Ministerio de
Educacion y Ciencia  under grant: BFM2003-00936


\end{document}